# Enhanced Self-Supervised Multi-Image Super-Resolution for Camera Array Images

Yating Chen, Feng Huang, Xianyu Wu, Jing Wu, Ying Shen

***Abstract*—**Conventional multi-image super-resolution (MISR) methods, such as burst and video SR, rely on sequential frames from a single camera. Consequently, they suffer from complex image degradation and severe occlusion, increasing the difficulty of accurate image restoration. In contrast, multi-aperture camera-array imaging captures spatially distributed views with sampling offsets forming a stable disk-like distribution, which enhances the non-redundancy of observed data. Existing MISR algorithms fail to fully exploit these unique properties. Supervised MISR methods tend to overfit the degradation patterns in training data, and current self-supervised learning (SSL) techniques struggle to recover fine-grained details. To address these issues, this paper thoroughly investigates the strengths, limitations and applicability boundaries of multi-image-to-single-image (Multi-to-Single) and multi-image-to-multi-image (Multi-to-Multi) SSL methods. We propose the Multi-to-Single-Guided Multi-to-Multi SSL framework that combines the advantages of Multi-to-Single and Multi-to-Multi to generate visually appealing and high-fidelity images rich in texture details. The Multi-to-Single-Guided Multi-to-Multi SSL framework provides a new paradigm for integrating deep neural network with classical physics-based variational methods. To enhance the ability of MISR network to recover high-frequency details from aliased artifacts, this paper proposes a novel camera-array SR network called dual Transformer suitable for SSL. Experiments on synthetic and real-world datasets demonstrate the superiority of the proposed method.

## I. Introduction

Due to the wide field of view, high temporal resolution, low distortion, and high sensitivity to moving objects, multi-aperture imaging system has been widely applied in biomedical [1], satellite remote sensing [2], motion detection [3], and robot navigation [4], among other fields. With the integration of micro-nano fabrication technology, multi-aperture imaging systems can achieve miniaturization [5]. However, these advantages come at the cost of sacrificing spatial resolution. Multi-aperture SR [5-7] can leverage the complementary aliased information from LR images with different apertures to reconstruct high-frequency details [5-8]. Furthermore, Multi-aperture SR overcomes the bottleneck of mutual constraints between image resolution and system track length in traditional single-aperture imaging systems, thereby reducing system thickness. Currently, LR images in deep learning-based SR methods [9-12] mainly come from mobile phones or the internet, and are therefore mainly applied in areas such as mobile photography or entertainment. In contrast, Multi-aperture SR with camera arrays is a computational optical imaging technology that jointly optimizes optical imaging systems and reconstruction algorithms. The motion between frames in both video and burst includes both overall camera motion and local object motion. Consequently, video and burst are susceptible to occlusion [11, 13], an inevitable and undesirable phenomenon. Furthermore, both video SR and burst SR encounter the issue of sampling configuration degradation. This means that the offsets among LR frames manifest as narrow elliptical distributions rather than a uniform distribution in a disk [14], thereby reducing stability and reliability in imaging quality. Due to the well-designed planar distribution and grid arrangement of camera array, camera-array SR is capable of avoiding the sampling degradation encountered in video SR and burst SR. Additionally, the compact structure and synchronized capture capabilities of camera array can effectively mitigate occlusion.

The current camera-array SR algorithms primarily adopt physical-based approaches, aiming to maximize posterior probability to estimate expected HR image [15-17]. However, these methods rely on explicit, handcrafted degradation kernels and priors, which cannot precisely capture the characteristics of real degradation kernels [18] and natural images [19]. Furthermore, physical-based approaches with sophisticated priors are often time-consuming to achieve better performance [19], such as sparse and low-rank priors. Currently, deep learning-based MISR methods are primarily designed for video and burst. Although burst SR methods can be adapted for multi-aperture SR, they still encounter challenges in recovering aliasing artifacts. These challenges include misinterpreting aliasing artifacts as genuine scene features, resulting in inconsistencies between the recovered image and the actual scene. Moreover, most of these methods rely on supervised training. The models trained via a supervised manner exhibit poor generalization ability. Supervised methods necessitate the collection of aligned LR-HR pairs, which can be impractical for every device and scene. In some specialized scenarios, such as remote sensing and medical imaging, obtaining HR ground-truth images can be challenging [14]. Unsupervised SR, on the other hand, only requires unpaired LR-HR data. However, most unsupervised learning is designed for single image super-resolution (SISR). Moreover, unsupervised learning, particularly with adversarial networks, can produce "hallucinated" high-frequency details

Yating Chen is with the Department of Precision Instruments, Tsinghua University, Beijing, 100084, China.

Feng Huang, Xianyu Wu, Jing Wu and Ying Shen are with the School of Mechanical Engineering and Automation, Fuzhou University, Fuzhou, 350108, China. E-mail: { huangf@fzu.edu.cn, xwu@fzu.edu.cn. }

Corresponding author: Feng Huang, Xianyu Wu

[20, 21]. The noise distribution in the real world is highly complex and challenging to model [22]. Whether supervised or unsupervised, these approaches also fail to leverage input image sequence-specific information during testing. Therefore, self-supervised SR emerges as a promising alternative, mainly encompassing two approaches: zero-shot and image-to-image. In zero-shot, training images are examples extracted solely from test image [23]. It's important to note that zero-shot inference time includes training time. The current image-to-image SR methods have two approaches: one is Multi-to-Single SSL, and the other is Multi-to-Multi SSL. In Multi-to-Single SSL, all LR images are fed into the network, and the LR reference image from the inputs is taken as the training target. The trained models tend to smooth out texture details [14]. The Multi-to-Multi SSL are further divided into two types: one is to use all LR images simultaneously as inputs and training targets [24], and the other is to divide all LR images into inputs and training targets [25]. The first type is prone to producing zipper edges, especially in scenes with large displacements. The second type sacrifices the number of input images during testing. The Multi-to-Multi SSL in this paper refers to the first type.

This paper investigates the network and learning strategy of camera-array SR. Based on the degradation model of camera-array SR, this paper designs the camera array super-resolution network featuring a dual self-attention Transformer (CASR-DSAT) for SSL. CASR-DSAT can more effectively decode aliasing artifacts and achieve content-agnostic SR imaging. The frequency separation based spatially adaptive Multi-to-Multi SSL loss suitable for camera arrays is proposed. And the strengths, weaknesses, and applicability boundaries of both Multi-to-Single SSL and Multi-to-Multi SSL are analyzed. To leverage the advantages of both SSL approaches, this paper proposes the Multi-to-Single-Guided Multi-to-Multi SSL framework. This framework provides a new paradigm for integrating deep neural network with classical physics-based variational methods. Existing paradigms typically involve alternately updating the target image constrained by regularization terms and the target image constrained by joint data and regularization terms during the optimization process, such as a plug-and-play approach. The proposed paradigm alternately updates the network parameters with regularization functionality and the network parameters with functionality of both data and regularization. Therefore, the output is directly obtained from the network with functionality of both data and regularization during testing, eliminating the need for multiple iterations, which enhances parallel computing efficiency and reduces runtime.

The contributions of this work can be summarized into four folds: 1) We conduct the analysis of the view disparity characteristics and statistical properties of camera array images, and propose CASR-DSAT to more effectively decode aliasing artifacts. 2) We investigate the strengths, weaknesses, and applicability boundaries of two SSL methods--an area that has not been explored in previous works, and propose a Multi-to-Single-Guided Multi-to-Multi SSL framework to leverage the advantages of both. This framework provides a new paradigm for integrating deep neural network with classical physics-based variational methods. 3) We construct two camera array imaging systems and use them to establish real-world datasets.

The source code, pre-trained models and datasets are available at https://github.com/luffy5511/CASR-DSAT.

## II. Related Work

### A. Real-world SR

[26] employed a high-order degradation with several repeated degradation processes to simulate real-world degradation. Recently, [27] proposed a degradation-aware self-attention-based Transformer model for learning the degradation representations with unknown noise. Subsequently, [28] proposed a degradation maps extractor to distinguish various degradations. [29] merged supervised pre-training with self-supervised learning. While significant results were achieved with real SISR, direct application to real MISR is impractical [30]. To employ real data for supervised training, some researchers have established real video and burst LR-HR datasets. This is primarily achieved through three methods: short-to-long focal length mobile phones or digital cameras, as well as mobile phone to digital camera [11, 31, 32]. However, the captured LR-HR pairs suffer from differences in brightness, color, and field of view. Correcting results for these differences significantly impacts the learning of network. Given the challenges in obtaining HR ground-truth images and the labor-intensive nature of creating LR-HR datasets, unsupervised learning is applied to real-world SR. Most unsupervised SR methods [21, 33] employ a two-step approach. First, degradation distribution is learned through adversarial training, and a generator produces pseudo-degraded LR images based on learned degradation distribution. Then, supervised learning is performed using these synthesized pseudo LR-HR pairs to train a SR network. Current unsupervised SR works are primarily centered on SISR and employs GAN training, which can easily synthesize "hallucinated" rather than genuine details. Moreover, the degradation distribution of real-world LR images is complex, making the transformation between two different domains challenging [34].

We observe that the traditional physical-based MISR methods rely solely on input LR image sequences and do not require external data to recover real details [8, 15-17]. This provides significant inspiration for SSL in MISR.

### B. Self-supervised Learning

**Self-supervised image denoising.** Recent research has shown that it is possible to train denoising networks even in the absence of clean images. Noise2Noise employed independent noisy image pairs as input and target to train denoising networks [35]. Dewil et al. [36] proposed multiframe-to-frame video denoising to ensure temporal consistency, using a dilated input stack to avoid equivalent mappings. In Noise2Void, with only a single noisy image available, the noisy image was decomposed into input and target pairs [37]. It employed a blind-spot masking scheme for training. Laine et al. [38] proposed a blind-spot network without masking to improve training efficiency. It utilized both the noisy measurement and the context of neighboring (noisy) pixels to jointly infer clean values, thereby enhancing denoising quality. Both Noise2Void and Laine's denoising approaches were based on the assumption that noise was spatially independent. To address spatially correlated noise while preserving detail, AP-BSN [39] introduced asymmetric pixel-shuffle downsampling and Li et al. [40] proposed spatially adaptive SSL.

**Self-supervised multi-image SR.** Nguyen et al. [14] proposed a deep shift-and-add MISR network that allows for an arbitrary number of bursts as input, and Multi-to-Single SSL strategy. This strategy selected one of the LR frames as the target and removes this target frame during training to achieve Noise2Noise denoising. By inputting a random number of images within a specified range during training, any number of burst SR can be achieved. Subsequently, Nguyen et al. applied the self-supervised strategy from [14] to multi-exposure SR tasks [41]. Due to the training target of the aforementioned SSL being solely the LR reference frame, the trained models tend to smooth out textures and details that have a low contrast relative to the noise level. Therefore, Nguyen et al. proposed Multi-to-Multi SSL [24], where the LR reference frame serves as the training target in flat regions, while in texture regions, all input frames serve as the training target. However, the models trained using this learning strategy tend to produce zipper edges, especially in scenes with large offsets. To utilize multiple frames as training targets and avoid generating zipper artifacts, Bhat et al. [25] divided the burst frames into input frames and training targets. However, this method sacrifices the number of input images during testing, which is more pronounced for devices like camera arrays with clearly defined input image number and for models where network parameters are related to the number of input images. Additionally, a significant disparity in the number of input images between training and testing can lead to a sharp decline in testing performance. Recently, Bai and Pan [42] proposed self-supervised blind video super-resolution. During training, the input of network includes LR reference frames, so relying solely on LR reference frames as training targets could lead to trivial solutions. Therefore, they added auxiliary constraints of zero-shot learning to regularize the blur kernel and HR frame estimation networks. Even though models can be fine-tuned using real datasets in real SR, the noise introduced in the auxiliary constraints remains simulated noise, hence the robustness to real-world noise needs to be further improved.

## III. Proposed Method

### A. Observation Model

The camera-array SR disentangles the aliasing artifacts in input LR images [43]. These artifacts emerge due to the inadequate sampling frequency of the detector, which fails to capture high-frequency signals transmitted through optical components. Consequently, high-frequency information in LR images is not lost but is encoded within low-frequency signals. The reconstruction of multiple aliased LR images with sub-pixel offsets allows the retrieval of high-frequency details [41, 44-47]. Random errors in the optical axis parallelism during camera array alignment and assembly ensure that the sampling deviation between apertures has a subpixel magnitude [48].

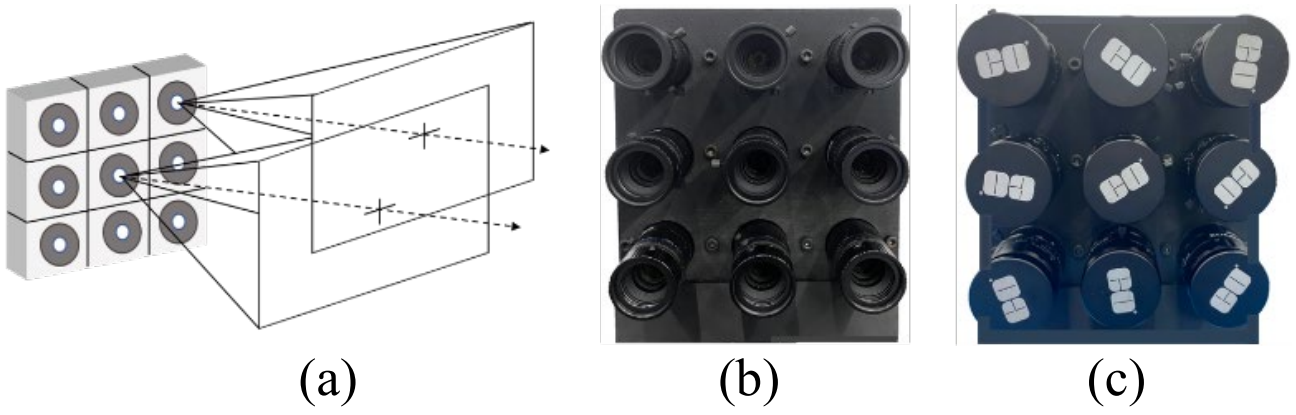

(a) (b) (c)

Fig. 1. (a) Imaging schematic diagram of camera-array SR. (b)The constructed nine-aperture camera-array SR imaging system using Basler lens. (c) The constructed nine-aperture camera-array SR imaging system using Edmund lens.

Fig. 1 represents imaging schematic diagram of the camera array system. The observation model of a camera array based SR imaging system can be expressed as follows:

$$\boldsymbol{I}_s^{LR} = \boldsymbol{D}_s \boldsymbol{W}_s \boldsymbol{I}^{HR} + \boldsymbol{e}_s \left( s = 1 \ldots N \right) \quad (1)$$

where $\boldsymbol{I}_s^{LR} \in \mathbb{R}^{C_{in} \times H \times W}$ represents the image captured by the $s$-th camera, $\boldsymbol{I}^{HR} \in \mathbb{R}^{C_{in} \times \sqrt{N}H \times \sqrt{N}W}$ represents the HR image to be estimated, $\boldsymbol{W}_s$ is the warping matrix defined by the geometrical transformation related to the image captured by the $s$-th camera with the reference image, $\boldsymbol{D}_s$ is the bi-dimensional downsampling operator for the imaging focal plane arrays which captures blur owing to pixel integration, $\boldsymbol{e}_s \in \mathbb{R}^{C_{in} \times H \times W}$ represents noise, $N$ is the number of apertures. Without considering factors such as registrational accuracy and optical aberration, the theoretical SR magnification for camera-array SR can be expressed as:

$$r_{SR} = \min\left( \frac{f_{diffraction}}{f_{Nyquist}}, \sqrt{N} \right) = \min\left( \frac{2AP}{1.22\lambda f}, \sqrt{N} \right) \quad (2)$$

where $f_{diffraction}$ is the diffraction limit angular frequency, $f_{Nyquist}$ is the Nyquist angular frequency, $P$ is the pixel size of detector, $A$ is the aperture diameter of the lens, and $\lambda$ is the wavelength. Fig. 1 (b) and Fig. 1 (c) depict the nine-aperture camera-array SR imaging systems constructed using the Basler lens and Edmund lens, respectively. Both systems are equipped with Hikvision cameras. For specific system parameters, refer to TABLE I. Compared to MISR, SISR benefits greatly from input images with large blur kernels. Because MISR methods can integrate information from multiple images, thus benefiting from clearer inputs (aliased LR images with smaller blur kernels) [49]. Fig. 2 illustrates the theoretical basis for applying SSL to camera-array SR.

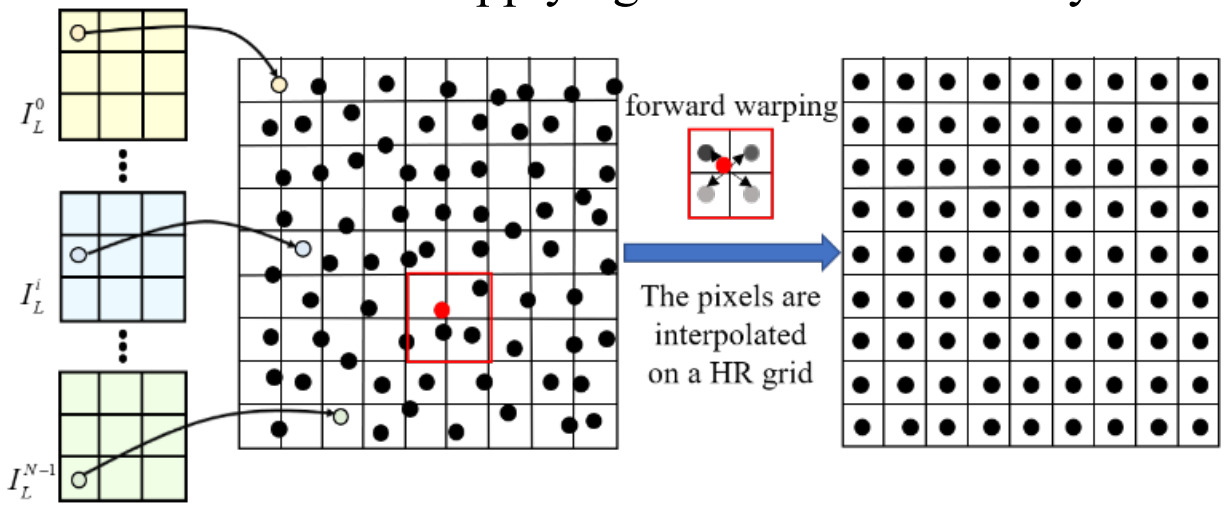

Fig. 2. The fundamental principle of SSL for camera-array SR imaging. Due to the parallax between apertures, the samples from different aperture images are located at different positions in HR image. By rearranging the samples from LR images, the sampling frequency can be increased. Therefore,

using only the input LR images as supervision signals is sufficient to train the SR network. Moreover, there is no need for clean images; independent noisy image pairs of the same scene are sufficient to train the denoising network.

TABLE I
PARAMETERS OF THE TWO CAMERA-ARRAY SR IMAGING SETUPS

| | |
|---|---|
| Lens type | Basler C125-2522-5M-P, Edmund 35 mm C visible–near-infrared |
| Camera type | Hikvision MV-CA004-10UM, Hikvision MV-CA004-10UM |
| Number of cameras | 9, 9 |
| Focal length | 25 mm, 35 mm |
| F-number | 2.2, 1.65 |
| Image sensor | Sony IMX287, Sony IMX287 |
| Sensor pixel size | 6.9 μm × 6.9 μm, 6.9 μm × 6.9 μm |
| Baseline length | 50 mm, 50 mm |
| $f_{\text{diffraction}}/f_{\text{Nyquist}}$ | 9.35, 12.5 |
| $r_{SR}$ | 3, 3 |

*B. Network Architecture*

The overview of the CASR-DSAT network is depicted in Fig. 3, comprising four modules: motion estimation, feature extraction, feature fusion, and feature reconstruction. CASR-DSAT employs FNet to calculate the aperture offsets, introduces the channel self-attention Transformer backbone (CSATB) to characterize the image, incorporates sub-pixel motion compensation (SPMC) to align image features, employs an adaptive fusion strategy of HR pseudo-camera array features, and introduces the spatial self-attention Transformer backbone (SSATB) for reconstructing fused features. Both self-attention Transformers have linear complexity with respect to the image size. More details of the network architecture can be found in the Supplementary Material.

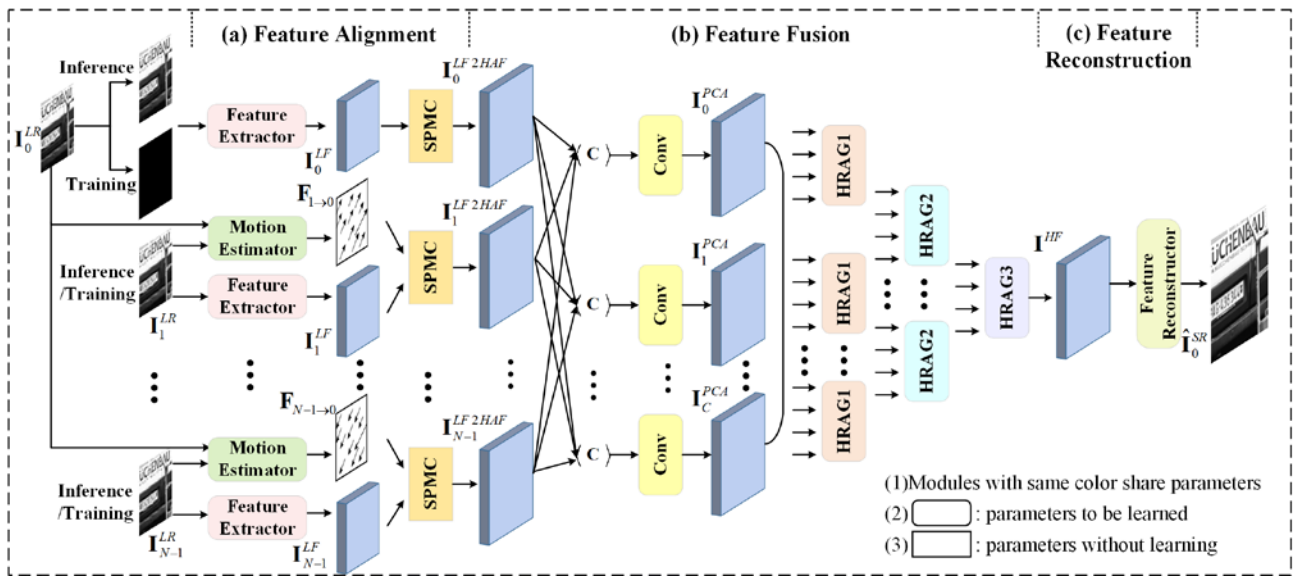

Fig. 3. The overview of the proposed CASR-DSAT network.

*C. Learning Strategy*

**Frequency separation based spatially adaptive Multi-to-Multi SSL.** The proposed Multi-to-Multi SSL takes $\boldsymbol{I}_0^{LR}$ as the training target for denoising in flat areas and all input LR images as the training target for detail enhancement in texture areas, thereby further enhancing fine-grained details and edge sharpness, as shown in Fig. 4 (a). Due to manufacturing tolerances and inter-aperture disparity, camera array images exhibit minor brightness variations. Directly using input images as training targets will result in SR images with speckle artifacts. Considering that $\boldsymbol{I}_0^{LR}$ as the training target ensures the restoration of low-frequency information, when using all input LR images as the training target, it is only necessary to focus on the recovery of high-frequency details. For this, high-frequency information of the input LR images is extracted as the training target. Flat areas and texture areas are determined based on the standard deviation. Because the standard deviation of high-frequency image is not influenced by lighting conditions, making it easier to distinguish between flat and textured regions, the high-frequency image is chosen instead of the intensity image to determine texture areas. The calculated texture confidence is as follows:

$$\boldsymbol{\alpha}_s(i,j)=\begin{cases} S\left(\boldsymbol{\sigma}_s(i,j)-\tau_\alpha^l\right), & \boldsymbol{\sigma}_s(i,j)<\tau_\alpha^l \\ 0.5, & \tau_\alpha^l \le \boldsymbol{\sigma}_s(i,j) \le \tau_\alpha^u \\ S\left(\boldsymbol{\sigma}_s(i,j)-\tau_\alpha^u\right), & \boldsymbol{\sigma}_s(i,j)>\tau_\alpha^u \end{cases} \tag{3}$$

where $\boldsymbol{\sigma}_s$ is the standard deviation of high-frequency image, $S(\ )$ is the Sigmoid function, $\tau_\alpha^l$ and $\tau_\alpha^u$ are constants. The proposed Multi-to-Multi self-supervised loss is,

$$\ell_{denoise}=\left\|\left(D\left(\hat{\boldsymbol{I}}_0^{SR}\right)-\boldsymbol{I}_0^{LR}\right)\odot\left(1-\boldsymbol{\alpha}_0\right)\right\|_1 \tag{4}$$

$$\ell_{detail}=\frac{1}{N}\sum_s\left\|\left(D\left(W\left(\hat{\boldsymbol{H}}_0^{SR},\sqrt{N}\boldsymbol{F}_{s\to 0}\right)\right)-\boldsymbol{H}_s^{LR}\right)\odot\boldsymbol{\alpha}_s\right\|_1 \tag{5}$$

$$\ell_{M2M}=\ell_{noise}+\ell_{detail}+\ell_{me} \tag{6}$$

where $W(\ )$ represents the backward warping operation based on the flow, $D(\ )$ represents the downsampling operation, $\ell_{me}$ represents the motion estimation loss from [14], $\hat{\boldsymbol{I}}_0^{SR}$ is the SR image outputted by the network, $\hat{\boldsymbol{H}}_0^{SR}$ and $\boldsymbol{H}_s^{LR}$ are respectively the high-frequency images of $\hat{\boldsymbol{I}}_0^{SR}$ and $\boldsymbol{I}_s^{LR}$.

**Multi-to-Single-Guided Multi-to-Multi SSL.** The Multi-to-Single SSL takes $\boldsymbol{I}_0^{LR}$ as the training target [14], as depicted in Fig. 4 (c). The models trained with Multi-to-Single SSL generate clean SR images with smooth edges and good visual perception. However, fine-grained details are prone to being smoothed out. In contrast, SR images generated by the models trained with Multi-to-Multi SSL can further enhance edge sharpness and detail recovery. However, they tend to produce zipper edges, especially in scenes with significant offsets. Traditional physics-based methods without image prior constraints produce zipper edges in SR images. Despite not having seen high-quality images, deep neural network reflects natural image priors in their network structures and parameter initializations, such as self-similarity and smoothness properties [63]. This trait makes deep neural network more adept at generating SR images with smooth edges. Hence, constraining the solution space and guiding network learning are crucial for addressing the limitations of Multi-to-Multi SSL. To address this, Multi-to-Single-Guided Multi-to-Multi SSL framework is proposed to fully leverage the advantages of both Multi-to-Single SSL and Multi-to-Multi SSL, as depicted in Fig. 4 (b). The edge-smoothing prior learned from Multi-to-Single is used to regularize the Multi-to-Multi SSL, resulting in clean images rich in details and with smooth edges. The Multi-to-Single-Guided Multi-to-Multi SSL framework consists of two branches. The first branch employs the Multi-to-Single self-supervised loss (Fig. 4 (c)) to guide the learning of the light green CASR-DSAT network. The second branch uses the SR image generated by the CASR-DSAT network from the first branch as a supervised target, which serves as a regularization term. The Multi-to-Multi self-supervised loss (Fig. 4 (a)), in combination with this regularization, constrains the reconstruction of large-scale textures in the deep green CASR-DSAT network, resulting in images with smooth edges and rich textures. The reconstruction of small-scale textures and fine details in the deep green CASR-DSAT network is solely constrained by the Multi-to-Multi self-supervised loss, producing fine-grained images. The parameters of the two CASR-DSAT networks are not shared and are independently optimized. During each optimization iteration, the parameters of the two CASR-DSAT networks are alternately updated. During testing, only the deep green CASR-DSAT network is used, and the SR image it outputs is taken as the final result.

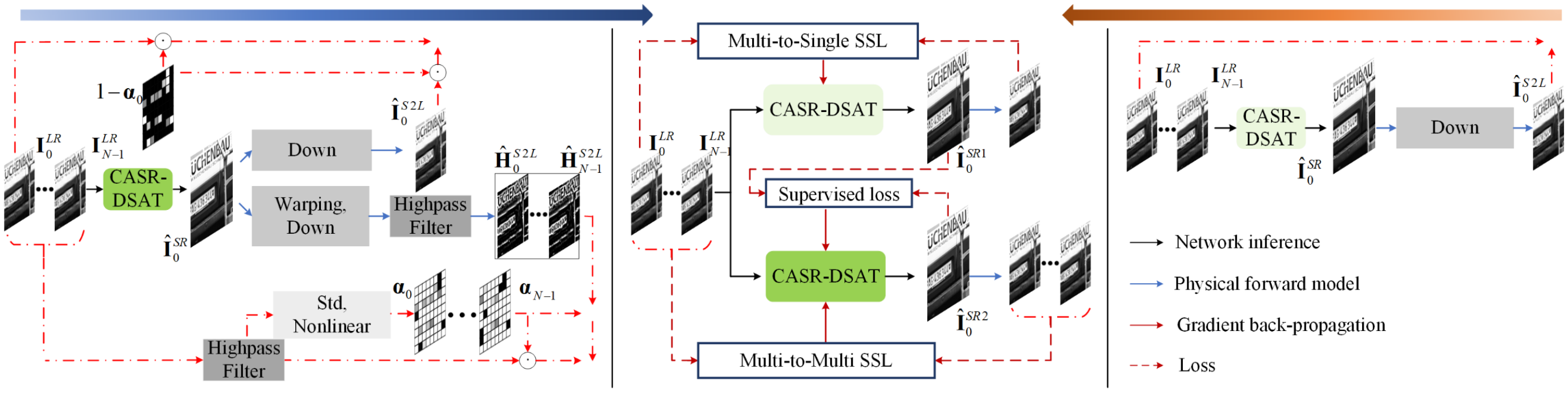

Fig. 4. The overview of the proposed Multi-to-Single-Guided Multi-to-Multi SSL framework. (a) is the Multi-to-Single SSL proposed by [14]. (c) and (b) are the Multi-to-Multi SSL and Multi-to-Single-Guided Multi-to-Multi SSL. In (b), the deep green CASR-DSAT network generates the SR image $\hat{I}_0^{SR2}$ as the final result. The SR image $\hat{I}_0^{SR1}$ generated by CASR-DSAT network trained with Multi-to-Single SSL, serves as an auxiliary supervised target, while the Multi-to-Multi self-supervised loss is used in conjunction to constrain the learning of the deep green CASR-DSAT network. The parameters of the two CASR-DSAT networks are not shared and are independently optimized in (b). During each optimization iteration, the parameters of the two CASR-DSAT networks are alternately updated.

The training loss function for the CASR-DSAT network in the first branch follows [14].The training loss function for the CASR-DSAT network in the second branch is as follows:

$$\ell_{GM2M} = \lambda \left\| \left( \hat{I}_0^{SR1} \text{-} \hat{I}_0^{SR2} \right) \odot U\left( \alpha_0 \right) \right\|_1 + \eta \ell_{M2M} \tag{7}$$

where $U(\cdot)$ denotes the upsampling operation, $\hat{I}_0^{SR1}$ and $\hat{I}_0^{SR2}$ represent the SR images outputted by the CASR-DSAT networks in the first and second branches respectively, $\lambda$ and $\eta$ set to 1.5 and 1.0, respectively. During testing, the CASR-DSAT network in the second branch is utilized.

Compared to directly using the model trained to completion for the first branch network, jointly training the two branch networks prevent the second branch network from exhibiting a bias towards SR images from Multi-to-Single SSL during training, thus mitigating the impact of Multi-to-Multi SSL.

## IV. Experiments

### A. Dataset and Implementation Details

For the synthetic training dataset, the LR image sequences are generated using the Zurich raw to RGB dataset[50] (46,239 images for training, 300 images for validation). The sRGB images are converted to grayscale images initially. Following [11], we generate a synthetic LR image sequence of size $N$ by applying random translations, random rotations (with ranges of [-24, 24] pixels and [-1°, 1°], respectively), ×3 downsampling and noise interference to HR grayscale image. Bilinear kernel is employed for image geometric transformation and downsampling. Following [14], the added noise is additive white Gaussian noise with a standard deviation of 3. The synthetic test data is also generated using the same operations. For training, each input crop is of size 56×56 pixels and the batch size is 15. Initially, we pretrain the motion estimation module FNet using the synthetic training dataset and the motion estimation loss from [14]. Subsequently, CASR-DSAT is trained end-to-end using Multi-to-Single self-supervised loss from [14] and the proposed Multi-to-Single-Guided Multi-to-Multi SSL, denoted as CASR-DSAT$_{M2S}$ and CASR-DSAT$_{GM2M}$ respectively. CASR-DSAT$_{M2S}$ and CASR-DSAT$_{GM2M}$ are trained using the synthetic training dataset and the pretrained FNet for 100 epochs. We utilize the Adam optimizer and cosine annealing scheduler [51] to train FNet for 50 epochs. And we use the AdamW optimizer [52], which is the common choice for training ViT models [53], along with the cosine annealing scheduler to train CASR-DSAT$_{M2S}$ and CASR-DSAT$_{GM2M}$. In the training of CASR-DSAT$_{M2S}$ and CASR-DSAT$_{GM2M}$, cosine annealing strategies for FNet and other modules is deployed for steadily decreasing the learning rate from $10^{-5}$ to $10^{-6}$ and from $10^{-4}$ to $10^{-6}$ respectively. In Multi-to-Single-Guided Multi-to-Multi SSL, $\tau_\alpha^l$ and $\tau_\alpha^u$ are respectively set to 8 and 15. The synthetic test data comprises publicly available datasets, including DIV2K[54] (100 natural scene images), PROBA-V [55] (100 satellite images), and the medical dataset [56] (50 MRI images). For the SR results on synthetic test data, the LR image sequences are generated following the steps of the synthetic training data.

Our system, as illustrated in Fig. 1 (b), captures outdoor natural scenes, yielding a total of 269 sets of camera array image sequences, with each set containing 9 LR images. Out of these sets, ten are randomly selected for the test dataset, while the rest are partitioned into 4,035 patches (4,025 for training and 10 for validation). Each input crop is of size 56×56 pixels. To evaluate the generalization ability of the proposed method across different scene domains, we employ the system in Fig. 1 (b) to capture ten indoor scenes for testing, involving books, dolls, food, resolution targets, and more. Additionally, to assess generalization in different systems with the same observation model, we employ the system in Fig. 1 (c) to capture the same indoor scenes. CASR-DSAT is fine-tuned using real training data for 25 epochs with the initial learning rate of $10^{-5}$. For the SR results on real test data, the LR image sequences are captured by the camera array system.

TABLE II
MISR RESULTS ON SYNTHETIC TEST DATASETS [54-56] FOR ×3 SR[a]

| Methods | DIV2K [54] | | | PROBA-V [55] | | | Medical Dataset [56] | | |
|---|---|---|---|---|---|---|---|---|---|
| | PSNR↑ | SSIM↑ | LPIPS↓ | PSNR↑ | SSIM↑ | LPIPS↓ | PSNR↑ | SSIM↑ | LPIPS↓ |
| Single Image | 23.16 | 0.661 | 0.205 | 27.28 | 0.694 | 0.148 | 20.75 | 0.621 | 0.160 |
| ML [48] | 28.16 | 0.848 | 0.103 | 32.65 | 0.892 | 0.067 | 25.89 | 0.856 | 0.084 |
| MAP-LR [15] | 29.12 | 0.858 | 0.111 | 33.47 | 0.886 | 0.089 | 26.96 | 0.858 | 0.088 |
| MAP-LRGSC [17] | 27.58 | 0.809 | 0.128 | 32.21 | 0.858 | 0.080 | 25.20 | 0.796 | 0.105 |
| DBSR [11] | 26.67 | 0.778 | 0.145 | 30.26 | 0.798 | 0.121 | 24.52 | 0.765 | 0.124 |
| MFIR [57] | 26.77 | 0.782 | 0.141 | 30.28 | 0.799 | 0.123 | 24.54 | 0.766 | 0.121 |
| BIPNet [58] | 27.02 | 0.787 | 0.134 | 30.35 | 0.801 | 0.116 | 26.63 | 0.768 | 0.121 |
| Burstormer [59] | 26.54 | 0.774 | 0.148 | 30.21 | 0.797 | 0.127 | 24.40 | 0.760 | 0.133 |
| DSA [14] | 19.23 | 0.660 | 0.168 | 21.88 | 0.650 | 0.163 | 17.14 | 0.623 | 0.168 |
| DSA* [14] | 28.86 | 0.862 | 0.082 | 33.48 | 0.902 | 0.055 | 27.24 | 0.876 | 0.060 |
| HDR-DSP [41] | 20.06 | 0.674 | 0.167 | 22.64 | 0.660 | 0.171 | 17.78 | 0.628 | 0.172 |
| HDR-DSP* [41] | 28.63 | 0.859 | 0.087 | 33.39 | 0.904 | 0.060 | 27.07 | 0.872 | 0.062 |
| CASR-DSAT$_{M2S}$ | **30.23** | **0.876** | **0.077** | **34.72** | **0.921** | **0.053** | **28.95** | **0.897** | **0.054** |
| CASR-DSAT$_{GM2M}$ | **29.92** | **0.867** | **0.070** | **34.46** | **0.919** | **0.036** | **28.95** | **0.896** | **0.045** |

[a]The red and green values represent the best and second best performances, respectively.

*B. SR Results on Synthetic Data*

This paper assesses the generalization capability of various methods across diverse datasets and compares deep learning-based and physics-based MISR methods. The physics-based MISR methods include ML [48], MAP-LR [15], and MAP-LRGSC [17]. For deep learning-based MISR methods, to ensure fair comparison, we retrain DBSR [11], MFIR [57], BIPNet [58], Burstormer [59], DSA [14], and HDR-DSP [41] using the same synthetic training dataset. The size of training crop remains 56×56 pixels, and the batch size is consistent with the original papers. Among these, DBSR, MFIR, BIPNet, and Burstormer are trained in a supervised manner, requiring ground-truth images. DSA and HDR-DSP are trained in a self-supervised manner and do not require ground-truth images. DSA and HDR-DSP use fixed blur kernels during training to enhance the sharpness of the output images. However, for simulated data, this blur kernel leads to oversharpening, resulting in DSA and HDR-DSP having the poorest metric values. Therefore, to more fairly compare the ability to restore details, DSA* and HDR-DSP* are introduced for comparison. DSA* and HDR-DSP* are unsharpened results of DSA and HDR-DSP training after removing the fixed blur kernel.

TABLE II illustrates the optimality of our CASR-DSAT. The suboptimal methods vary across different datasets. However, across all datasets, both CASR-DSAT$_{M2S}$ and CASR-DSAT$_{GM2M}$ achieve PSNR gain of more than 1.0 dB over the suboptimal methods, confirming the generalization capability of our approach. As shown in TABLE II, CASR-DSAT$_{M2S}$ has higher PSNR, while CASR-DSAT$_{GM2M}$ achieves superior LPIPS. CASR-DSAT$_{M2S}$ tends to generate smooth images, while CASR-DSAT$_{GM2M}$ focuses more on the texture structure of the scene.

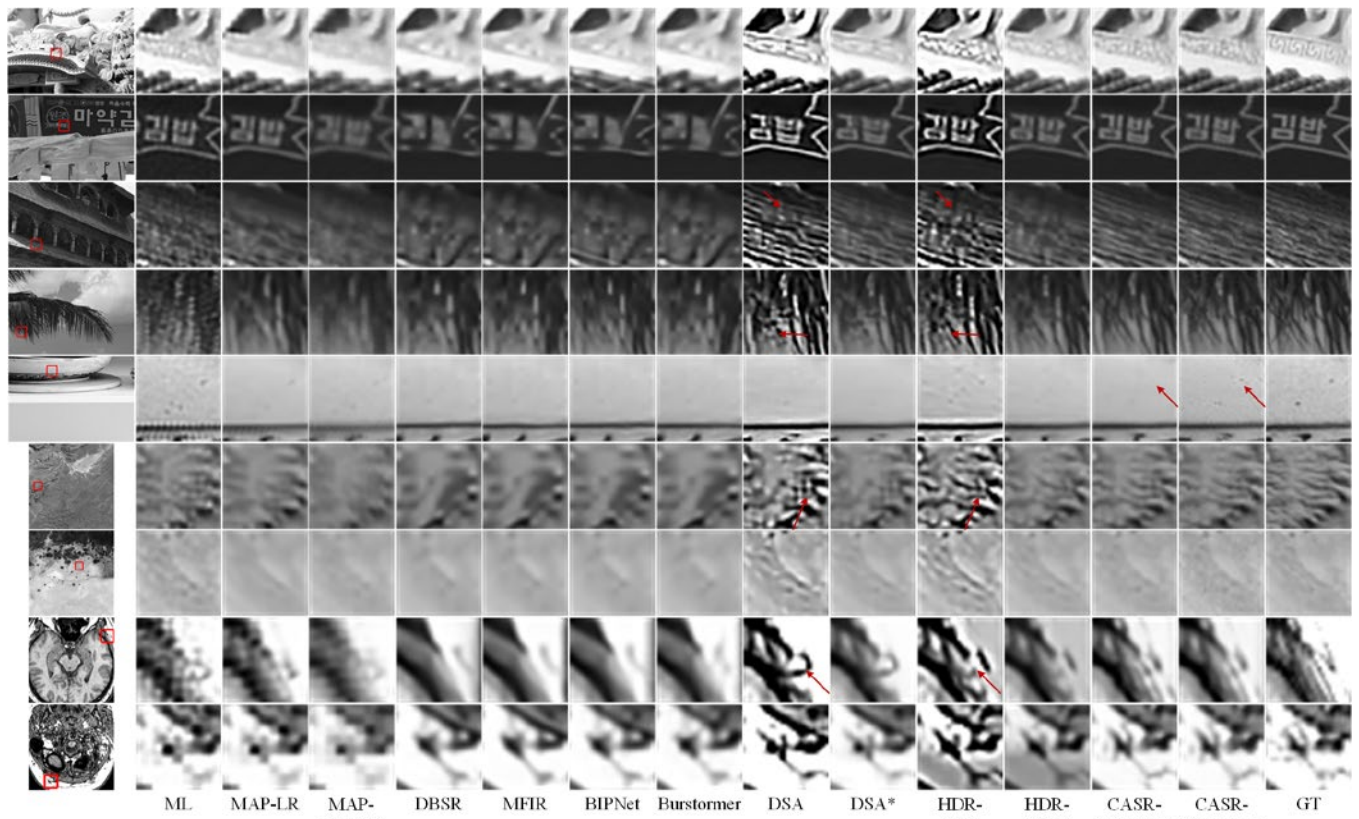

Fig. 5. Comparisons for ×3 MISR on synthetic DIV2K [54] (natural scene images), PROBA-V [55] (satellite images), and the medical dataset. Our CASR-DSAT$_{M2S}$ and CASR-DSAT$_{GM2M}$ restore fine texture and not hallucinates details. DSAT$_{GM2M}$ further enhances detail restoration, especially for low-contrast fine-grained details.

Traditional motion estimation methods perform inferiorly on small-sized images compared to DL, such as the medical dataset, resulting in a decline in the performance of physics-based SR methods, as verified in [60]. However, even in the perfectly aligned DIV2K dataset, our method surpasses physics-based methods. The advantages of SSL are highlighted when evaluating in scene domains different from the training dataset. Fig. 5 indicates that our CASR-DSAT is more effective in decoding aliasing artifacts, restoring fine textures and details that are more faithful to the ground truth. Compared to CASR-DSAT$_{M2S}$, CASR-DSAT$_{GM2M}$ can recover textures and details that have a low contrast relative to the noise level, as shown in the 5th and 7th rows of Fig. 5. ML exhibits zipper edges and noise artifacts. MAP-LR and MAP-LRGSC exhibit zipper artifacts along unaligned edges and smooth out fine details. Supervised learning methods can only recover large-scale contours. DSA and HDR-DSP are oversharpening, leading to ringing artifacts. DSA* and HDR-DSP* struggle to recover fine textures and may produce details inconsistent with the ground truth.

### C. SR Results on Real Data

As self-supervised methods do not require ground-truth images, we fine-tune DSA and HDR-DSP using real training dataset for 25 epochs with the initial learning rate of $10^{-5}$. For the degradation of real data, we employ the sharpening training from [14] to improve visual perception of DSA, HDR-DSP and our CASR-DSAT. Because DBSR, MFIR, BIPNet, and Burstormer cannot be trained on real dataset without ground-truth images, they utilize the models pre-trained on synthetic data to test the real data. The indoor test scenes or similar ones do not appear in the training data. The testing results of the two CASR imaging systems utilize the model fine-tuned on the real training dataset created by the first CASR imaging system.

Fig. 6 demonstrates the real SR generalization ability of our method across scenes and systems. Compared to other DL methods, our CASR-DSAT can accurately restore fine textures and details, as seen in the stripes in the first row of Fig. 6. Compared to other physics-based methods, our CASR-DSAT exhibits superior visual results. According to TABLE III, it can be observed that our CASR-DSAT$_{M2S}$ achieves the superior MUSIQ. SR images generated by CASR-DSAT$_{GM2M}$ exhibit low sharpness in some test scenes due to individual aperture defocusing blur, resulting in poor performance in NIQE and MUSIQ. It is worth noting that metrics like NIQE and MUSIQ, although they can to some extent reflect the sharpness and texture richness of images, cannot accurately assess whether reconstructed details are faithful to real scenes. Furthermore, in the second systems of TABLE III, contradictory conclusions are observed between NIQE and MUSIQ for MAP-LRGSC. Therefore, we supplement quantitative metrics for SR magnification. The numerical values in Fig. 7 represent SR magnification [61] determined by MTF based on the Siemens star image. The proposed CASR-DSAT achieves a SR magnification surpassed only by ML (physics-based method). While ML has a higher SR magnification, the reconstructed images suffer from severe noise and exhibit aliasing artifacts at edges.

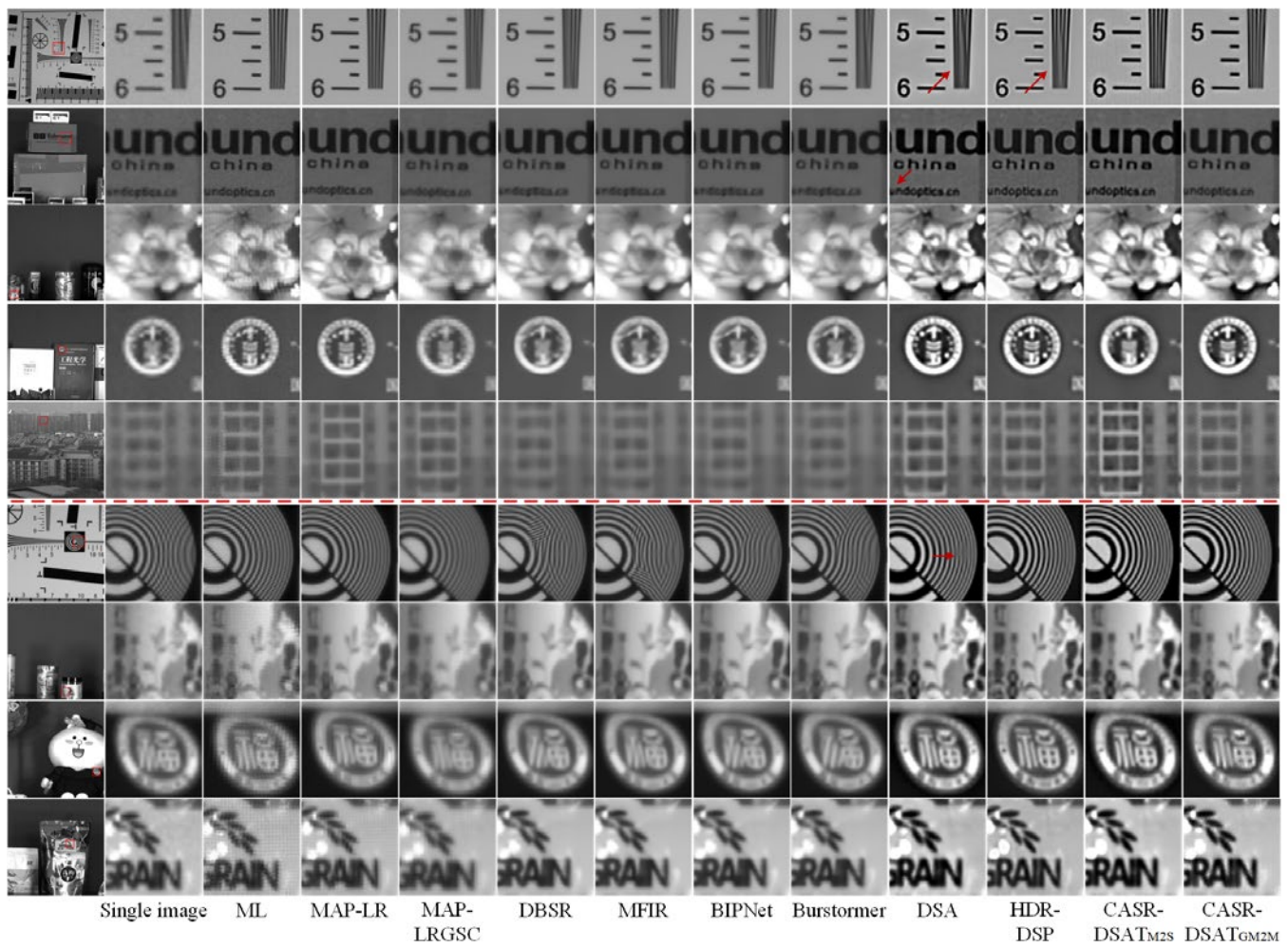

Fig. 6. Camera-array SR results (×3) on real camera array test dataset. Above and below the red dashed line are the data captured by the first and second camera-array SR imaging systems, respectively.

TABLE III

RESULTS OF ×3 CASR ON THE REAL CAMERA ARRAY TEST DATASETS[b]

| Methods | CASR1 | | CASR2 | |
|---|---|---|---|---|
| | NIQE ↓ | MUSIQ ↑ | NIQE ↓ | MUSIQ ↑ |
| Single Image | 5.48 | 34.77 | 5.42 | 36.25 |
| ML | **4.82** | 38.86 | 6.51 | 33.27 |
| MAP-LR | **4.61** | 45.05 | **4.97** | 41.48 |
| MAP-LRGSC | 5.23 | 34.39 | **5.34** | 32.44 |
| DBSR | 5.98 | 38.34 | 5.70 | 37.46 |
| MFIR | 6.05 | 38.88 | 5.75 | 37.61 |
| BIPNet | 5.99 | 38.25 | 5.66 | 37.37 |
| Burstormer | 6.02 | 38.73 | 5.57 | 37.88 |
| DSA | 5.04 | 46.75 | 5.38 | 44.19 |
| HDR-DSP | 4.98 | **48.53** | 5.43 | **44.59** |
| CASR-DSAT$_M$ | 5.11 | **47.66** | 5.47 | **44.82** |
| CASR-DSAT$_G$ | 5.44 | 36.15 | 5.75 | 35.82 |

[b] CASR-DSAT$_M$: CASR-DSAT$_{M2M}$; CASR-DSAT$_G$: CASR-DSAT$_{GM2M}$

## V. ANALYSIS AND DISCUSSIONS

### A. Ablation Study

**Importance of network components.** The ablation experiments are conducted on the three key components of the CASR-DSAT network: the feature extraction module CSATB, the feature fusion module HRPAF, and the feature reconstruction module SSATB. The baseline model employs the encoder and decoder from [14] as feature extraction and feature reconstruction modules. They are composed of a stack of several residual blocks. The motion estimation module employs FNet [49], and feature alignment is achieved using backward warping. A simple concatenation operation is used for fusion, and pixel shuffle is employed for upsampling. As shown in TABLE IV, A1-A7 indicate that the specified modules, marked with checkmarks, replace the corresponding modules in the baseline, while the remaining modules remain unchanged. For example, A1 indicates that the checked CSATB module replaces the feature extraction module in the baseline. A1 to A7 are trained for 30 epochs on the synthetic training dataset described in Section IV-A. The trained models are then tested on the synthetic test dataset, DIV2K, also described in Section IV-A, with the PSNR results presented in TABLE IV. When substituting SSATB for the decoder of the baseline, the grayscale distortion occurs, resulting in a significant decrease in PSNR. Except for A3, adding the proposed module to the baseline improves PSNR. The more modules added, the greater the improvement. Ultimately, our CASR-DSAT outperforms the baseline by 13.68 dB.

### B. CASR-DSAT: The Superiority of SSL

In the original paper, DBSR, MFIR, BIPNet, and Burstormer are trained using supervised learning. Because the Multi-to-Single SSL framework in [14] is highly versatile and can be adapted to DBSR, MFIR, BIPNet, and Burstormer networks. Therefore, to validate the superiority of the proposed CASR-DSAT network for SSL, we employ the Multi-to-Single SSL to train DBSR[11]、MFIR[57]、BIPNet[58]、Burstormer[59] networks. The size of training parameters are consistent with the original papers. In the M2S self-supervised training of DBSR and MFIR, $\boldsymbol{I}_0^{LR}$ are only used to calculate optical flow and the weighting values for fusion, but it itself does not participate in feature fusion. Because BIPNet and Burstormer networks utilize implicit feature alignment with edge enrichment and reference-based feature enrichment respectively, reference image features are also integrated during feature alignment. Therefore, $\boldsymbol{I}_1^{LR}$ is used as the reference image to align the other images, while $\boldsymbol{I}_0^{LR}$ is not inputted into the network. As shown in the first row of Fig. 8, the BIPNet network completely loses scene information. The DBSR network only reconstructs rough scene outlines. The MFIR network fails to recover high-frequency details. The SR images generated by the Burstormer network exhibit grayscale distortion. The first row of TABLE V displays their PSNR.

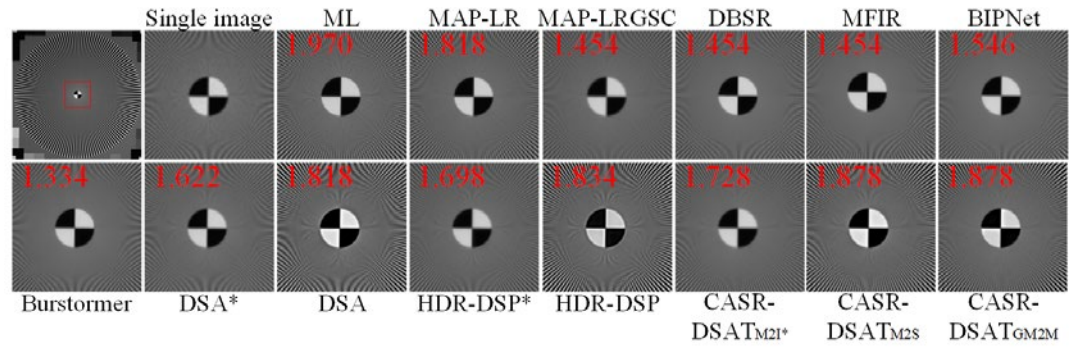

Fig. 7. Camera-array SR results (×3) on real Siemens star.

TABLE IV

IMPORTANCE OF CASR-DSAT COMPONENTS TRAINED FOR 30 EPOCHS WITH MULTI-TO-SINGLE SSL EVALUATED ON SYNTHETIC DIV2K [54] TEST DATASET

| | CSATB | HRPAF | SSATB | PSNR↑ |
|---|---|---|---|---|
| Baseline | | | | 15.63 |
| A1 | ✓ | | | 17.07 |
| A2 | | ✓ | | 15.96 |
| A3 | | | ✓ | 8.93 |
| A4 | ✓ | ✓ | | 22.47 |
| A5 | ✓ | | ✓ | 17.78 |
| A6 | | ✓ | ✓ | 25.79 |
| A7 | ✓ | ✓ | ✓ | **29.31** |

### *C. M2S, M2M, and GM2M Comparative Analysis*

Train the DBSR, MFIR, BIPNet, Burstormer, DSA, HDR-DSP, and CASR-DSAT networks using Multi-to-Multi SSL from Section III-C. Training epochs for DBSR, MFIR, BIPNet, Burstormer, DSA, and HDR-DSP are halved from their original papers, while CASR-DSAT training epochs are halved from M2S training. To balance denoising and detail enhancement, $\tau_\alpha^l$ and $\tau_\alpha^u$ in M2M are set to 20 and 30, respectively. During Multi-to-Multi SSL, both training and testing utilize complete 9 LR images, with $\boldsymbol{I}_0^{LR}$ engaged throughout the network process. Since BIPNet and Burstormer adopt implicit alignment, no optical flow estimation is needed. To compute the error between the images after geometric deformation and downsampling of $\hat{\boldsymbol{I}}_0^{SR}$ and $\left\{\boldsymbol{I}_s^{LR}\right\}_{s=0}^{N-1}$, a pre-trained FNet [49] is used to estimate the optical flow from $\left\{\boldsymbol{I}_s^{LR}\right\}_{s=1}^{N-1}$ to $\boldsymbol{I}_0^{LR}$.

**Limitations of M2M.** Due to the issue of reconstructed images being biased towards darkness in some networks during the Multi-to-Multi SSL in Section III-C, to explore the intrinsic problems of Multi-to-Multi SSL and since synthetic data does not have the brightness difference between LR images, grayscale images are used instead of high-frequency images to retrain the DBSR, MFIR, BIPNet, Burstormer, DSA, and HDR-DSP networks. The results of the two Multi-to-Multi SSL methods are denoted as $M2M_F$ and $M2M_G$, respectively. The results of DBSR, MFIR, BIPNet, and Burstormer trained using Multi-to-Multi SSL, as shown in the second and third rows of Fig. 8 and TABLE V, are significantly different from the ground truth.

TABLE V

PSNR OF DIFFERENT SUPERVISED LEARNING NETWORKS WITH MULTI-TO-SINGLE SSL AND MULTI-TO-MULTI SSL ON SYNTHETIC DIV2K [54] TEST DATASET

| Methods | DBSR | MFIR | BIPNet | Burstormer |
|---|---|---|---|---|
| M2S | 13.56 | 19.50 | 4.66 | 12.11 |
| $M2M_F$ | 14.72 | 13.25 | 6.80 | 8.18 |
| $M2M_G$ | 14.03 | 13.57 | 15.17 | 15.12 |

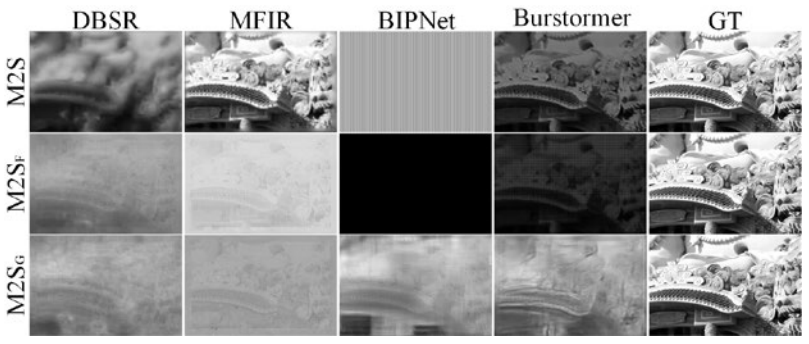

Fig. 8. Results of DBSR, MFIR, BIPNet, Burstormer with Multi-to-Single SSL and Multi-to-Multi SSL.

To analyze the limitations of applying Multi-to-Multi SSL to DSA, HDR-DSP, and our CASR-DSAT network, we conducted tests with two different offset ranges. Since the results of $M2M_F$ and $M2M_G$ are essentially the same, only one set of results is presented. The DIV2K synthetic datasets are generated using two different offset ranges: [-24, 24] pixels and [-3, 3] pixels, with rotation angles ranging from [-1°, 1°]. The offset range of [-24, 24] is derived from [11] and serves as a widely recognized benchmark. Meanwhile, the offset range of [-3, 3] is based on [14]. For ×3 SR, [-3, 3] achieves sub-pixel offsets. The PSNR for both offsets in TABLE VI is calculated within the same region, namely the common overlapping area of image sequences under the offset of 24 pixels. As indicated in TABLE VI, the PSNR for small offsets is approximately 1.0 dB higher than for large offsets. For scenes with large offset motion, DSA, HDR-DSP, and CASR-DSAT networks exhibit severe zipper artifacts, as evident in Fig. 9.

**Effectiveness of GM2M.** Observing Fig. 10, it can be noted that the proposed Multi-to-Single-Guided Multi-to-Multi SSL framework leverages the advantages of both Multi-to-Single SSL and Multi-to-Multi SSL. It not only restores fine-grained texture but also eliminates the zipper artifacts present in Multi-to-Multi SSL. In DIV2K, PROBA-V, and the medical dataset, PSNR of Multi-to-Single-Guided Multi-to-Multi SSL is higher than that of Multi-to-Multi SSL by 3.94, 3.71, and 5.34 dB, respectively.

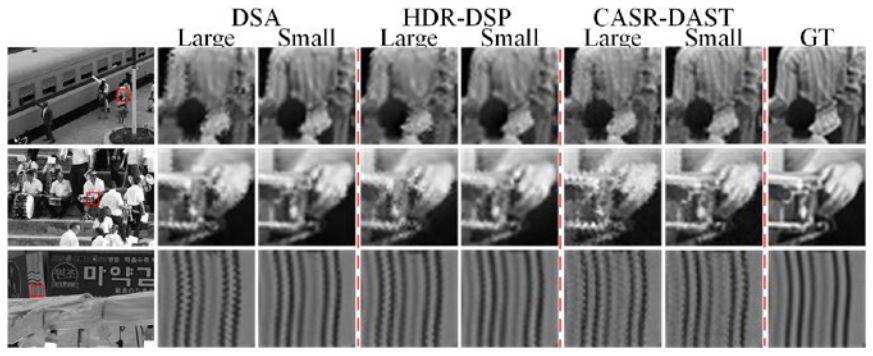

Fig. 9. Results of DSA, and HDR-DSP with Multi-to-Multi SSL.

TABLE VI
PSNR of DIFFERENT SSL NETWORKS with Multi-to-Multi SSL on SYNTHETIC DIV2K [54] TEST DATASET

| Methods | DSA | HDR-DSP | CASR-DSAT |
|---|---|---|---|
| Large offset | 28.25 | 28.15 | 25.98 |
| Small offset | 29.56 | 29.35 | 26.84 |

### *D. Model Size, Running Time, and Computational Complexity Comparisons*

On the DIV2K synthetic dataset where the size of the input LR image sequence is 112×168×9, the runtime of different algorithms and the FLOPs and model size of DL methods are shown in TABLE VII. The physics-based methods are executed on MATLAB software running on an Intel(R) Xeon(R) W-2235 processor (3.80 GHz) with 128 GB of memory. DL methods are implemented using NVIDIA A100 80GB PCIe GPU. The first three methods in TABLE VII are traditional physics-based model methods, so their parameter counts and FLOPs are not reported. From TABLE VII, it can be seen that among all methods, the FLOPs and runtime of the proposed CASR-DSAT network rank in the middle. Although the parameter size of the proposed network is higher than other methods, compared to AFCNet [62], which also uses channel self-attention Transformer blocks [63] and has a parameter size of 47M, the parameter size of our network is only 14.45M. This reflects that the performance improvement of our method is not due to using large models, but rather stems from adopting the architecture of the dual self-attention Transformer.

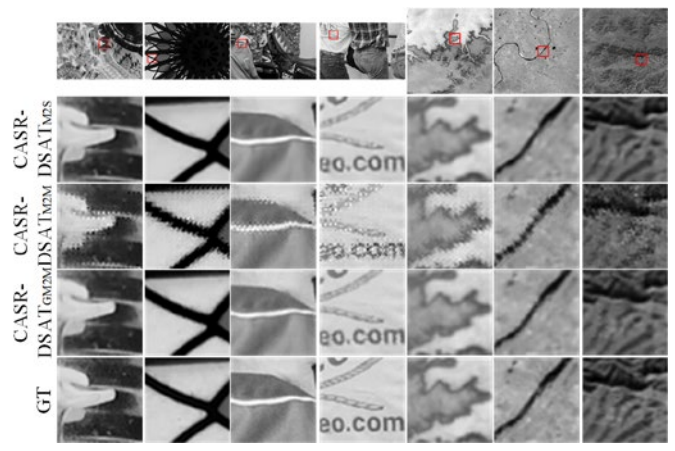

Fig. 10. Comparison of CASR-DSAT networks trained using three SSL methods (M2S, M2M, and GM2M).

TABLE VII
MODEL SIZE, RUNNING TIME, AND COMPUTATIONAL COMPLEXITY COMPARISONS

| Methods | ML | MAP-LR | MAP-LRGSC | DBSR | MFIR |
|---|---|---|---|---|---|
| Parameters(M) | | | | 13.01 | 12.13 |
| Running time(/s) | 1.39 | 51.77 | 184.33 | 0.34 | 0.38 |
| FLOPs(G) | | | | 529.41 | 660.17 |
| **Methods** | **BIPNet** | **Bursto-rmer** | **DSA** | **HDR-DSP** | **CASR-DSAT** |
| Parameters(M) | 6.67 | 3.58 | 2.85 | 2.85 | 14.45 |
| Running time(/s) | 0.91 | 0.20 | 0.24 | 0.40 | 0.49 |
| FLOPs(G) | 2442.63 | 191.90 | 262.88 | 275.46 | 558.86 |

## VI. CONCLUSION

This study explores a novel MISR technique, that is, camera-array SR technology based on SSL. On one hand, it addresses issues encountered in video SR and burst SR, such as sampling degradation and severe occlusion effects. On the other hand, it overcomes the trade-off between runtime and reconstruction quality inherent in traditional physics-based SR algorithms. Based on the statistical characteristics of observed images, we propose a CASR-DSAT network suitable for SSL aimed at enhancing detail recovery. We analyze the strengths, limitations and applicability boundaries of Multi-to-Single SSL and Multi-to-Multi SSL and further propose Multi-to-Single-Guided Multi-to-Multi SSL that combines the advantages of both SSL methods. It provides a new paradigm for combining deep neural network with classical physics-based variational methods. Compared to plug-and-play approach that updates the target image during iterations, it updates network parameters iteratively, thereby eliminating the need for iterations during testing and improving operational efficiency. In multiple simulated datasets across various scene domains, the proposed method achieves a PSNR improvement of at least 1.0 dB over state-of-the-art methods. In real camera array datasets, the proposed method accurately restores texture and detail, with SR magnification exceeding that of state-of-the-art deep learning methods. The FNet used in this paper is effective for motion compensation in most scenarios. However, as FNet is a fixed-kernel, fully convolutional optical flow estimation network, it may struggle with explicitly compensating for complex occlusions in real-world scenes. This is a limitation of the current approach. To address this, we plan to explore the use of spatially adaptive deformable convolutions, which offer more flexibility for motion compensation, in future work.

# Enhanced Self-Supervised Multi-Image Super-Resolution for Camera Array Images

## Supplementary Material

### *1. Network architecture*

Here, we provide additional details about our camera array super-resolution network architecture.

**Motion estimator and compensation with camera array images.** To ensure the range and accuracy of depth estimation, the baseline distance between the apertures in a stereo system is large, which results in significant disparity and introduces severe occlusion effects. The structure of camera array system is compact, so compared to a stereo system, the inter-aperture view disparity is smaller and occlusion effect is weaker. Implicit alignment is more effective in handling occlusion issues than explicit alignment. Due to the weak occlusion effect in camera array images and to maximize the preservation of sub-pixel information during alignment resampling operations, we opted for an explicit alignment method capable of achieving sub-pixel motion compensation. The inter-frame offsets in burst and video stem from both global camera motion and scene variations [1]. Only camera motion exists in the camera array. In conclusion, the occlusion effect in camera array is weak. Therefore, after balancing imaging performance and processing speed, simple optical flow network is used for motion compensation, which works well for most scenarios. Taking the central aperture image as the reference image, FNet [2] is used to estimate the offset of other aperture images to $\boldsymbol{I}_0^{LR}$,

$$\boldsymbol{F}_{s\to 0} = \mathrm{FNet}\left(\boldsymbol{I}_s^{LR}, \boldsymbol{I}_0^{LR}; \boldsymbol{\theta}_{FNet}\right) \in \mathbb{R}^{2\times H\times W} \tag{1}$$

The SPMC layer from [3] is employed to achieve sub-pixel motion compensation. SPMC utilizes a forward warping structure. Fig. 1 illustrates the difference between forward warping and backward warping. Specifically, SPMC maps the pixels of the LR image to a grid on the HR reference image based on the computed offset, as shown by the blue dot on the HR reference image in Fig. 1 (b). The blue dot is interpolated onto the neighboring grid points based on the weights of the bilinear sampling kernel. Subsequently, it performs weighted aggregation of pixel values on the grid points, resulting in the HR aligned image shown in Fig. 1 (b). This motion compensation approach preserves sub-pixel information between apertures and the sparsity in mapping to HR space. It avoids compromising inter-image complementary information and prevents the interpolated pixels in upsampling from dominating during training.

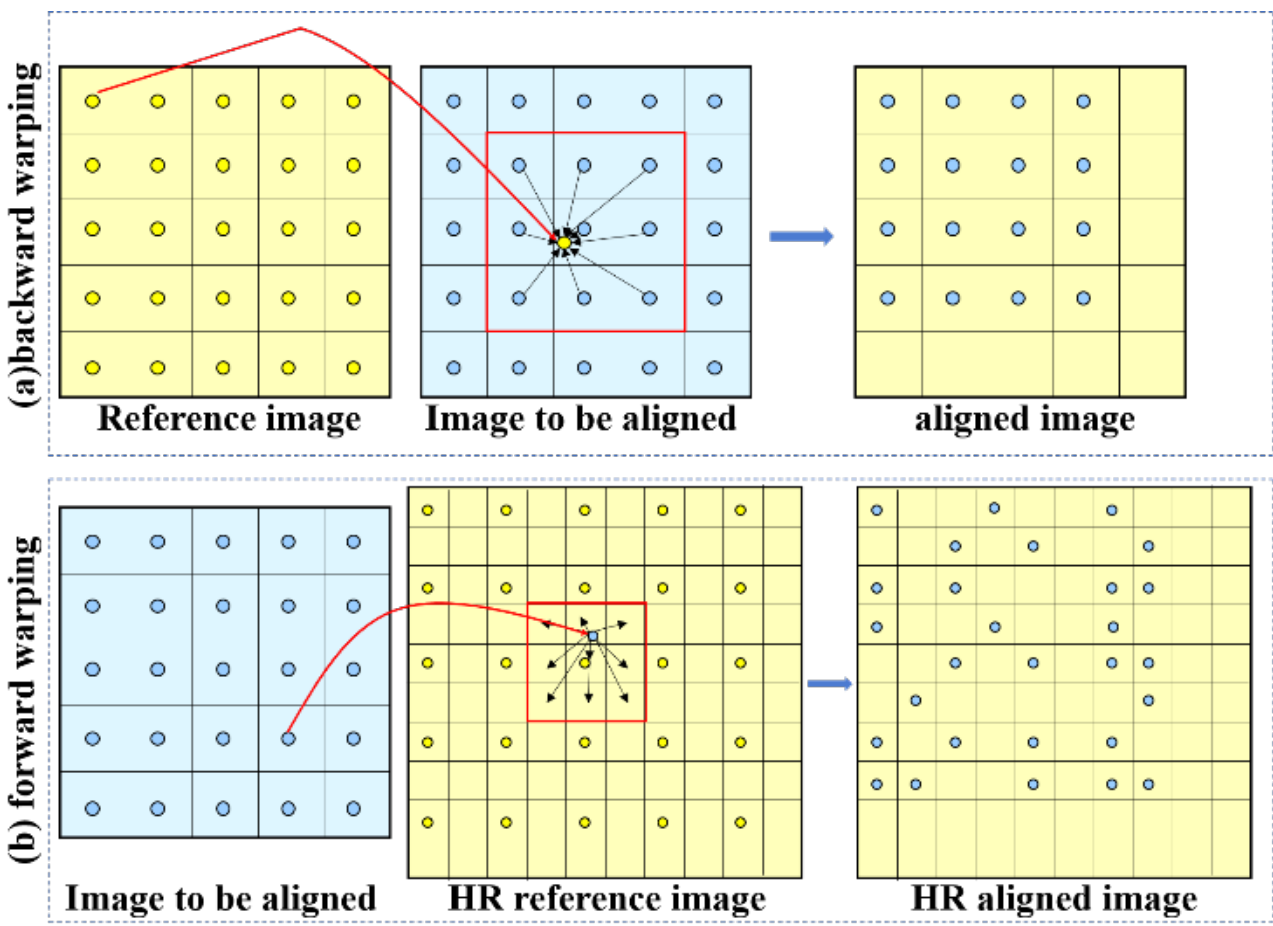

Fig. 1 Comparison between forward warping and backward warping.

**Channel self-Attention Transformer backbone.** Compared to convolutional neural network (CNN), Vision Transformer benefits from its self-attention mechanism and long-range dependency, meaning that it can easily derive global information and less need for vision-specific inductive bias [4]. Restormer [5] uses Transformer block with self-attention between feature channels rather than tokens for HR image restoration. Contextual information from both the pixel space and channel dimensions can enhance the feature representation [6]. Therefore, we fine-tune the Restormer architecture for feature representation, which is termed the Channel Self-Attention Transformer Backbone (CSATB) in Fig. 2. On the one hand, channel self-attention can preserve complementary aliasing information in spatial dimensions, and on the other hand, it can enhance the synergy between the feature extraction and feature fusion modules. CSATB adopts a U-net structure to achieve multi-scale feature representation, where the aggregation of encoder features and decoder features is beneficial for preserving both structural and texture information. The residual connections at each level enhances training stability. Additionally, the convolution in shallow level of CSATB and Transformer blocks helps to extract local information. As the fusion module receives HR features as inputs, we choose CSATB over shallow feature extraction network. This choice allows us to use the fusion module with smaller parameter without compromising SR performance. The feature extracted by CSATB is as follows:

$$\boldsymbol{I}_s^{LF} = \mathrm{CSATB}\left(\boldsymbol{I}_s^{LR}; \boldsymbol{\theta}_{CSATB}\right) \in \mathbb{R}^{C\times H\times W} \tag{2}$$

The aligned feature after SPMC is as follows:

$$\boldsymbol{I}_s^{LF2HAF} = \mathrm{SPMC}\left(\boldsymbol{I}_s^{LF}, \boldsymbol{F}_{s\to 0}\right) \in \mathbb{R}^{C\times\sqrt{N}H\times\sqrt{N}W} \tag{3}$$

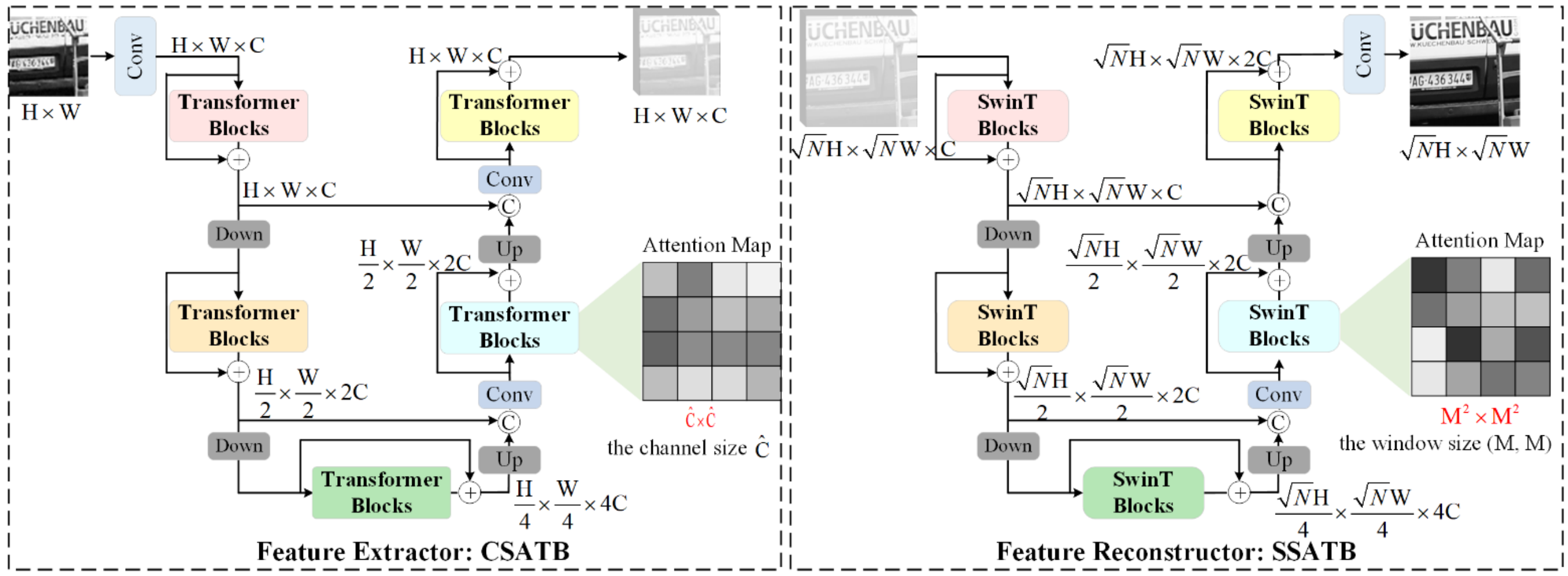

Fig. 2. The overview of the feature extractor CSATB and the feature reconstructor SSATB in the CASR-DSAT network. In CSATB and SSATB, the number of Transformer blocks at three levels are [1, 2, 4] and [2, 2, 4], corresponding attention heads are [1, 2, 2] and [2, 2, 4], and the number of channels is [64, 128, 256] in both modules.

Inspired by [7], we design the HR pseudo-camera-array adaptive fusion module (HRPAF) for feature fusion. $\left\{\boldsymbol{I}_s^{LF2HAF}\right\}_{s=0}^{N\text{-}1}$ are rearranged, i.e., its corresponding channel-wise features are concatenated to generate pseudo-camera-array. And every four pseudo-camera-array features $\boldsymbol{I}_c^{PCA}$ are partitioned into a group sequentially. In each group, $\boldsymbol{I}_c^{PCA}$ and associated attention maps [7] are element-wise multiplied to achieve adaptive fusion. In comparison to the global maximum pooling as in [8] or the joint pooling as in [9], this fusion method can achieve adaptive fusion based on the image content and noise level. It performs average fusion on noise and edge-aware fusion on edges. The three-layer progressive fusion structure employed is better suited for sparse HR aligned features, enabling more effective aggregation of complementary information. The fused HR feature is as follows:

$$\boldsymbol{I}^{HF} = \text{HRPAF}\left(\left\{\boldsymbol{I}_s^{LF2HAF}\right\}_{s=0}^{N\text{-}1}; \boldsymbol{\theta}_{\text{HRPAF}}\right) \in \mathbb{R}^{C\times\sqrt{N}H\times\sqrt{N}W} \tag{4}$$

**Spatial self-attention Transformer backbone.** The fused features enable inter-aperture communication and recover high-frequency details. However, due to the sparsity of $\boldsymbol{I}_s^{LF2HAF}$ (containing many zero values) and residual noise interference, CNN with limited receptive field cannot perfectly remove residual noise and smooth edges for feature reconstruction [10]. To address this issue, the Swin Transformer block [11] with spatial self-attention is employed to replace the channel self-attention Transformer in CSATB, which is termed the Spatial Self-Attention Transformer Backbone (SSATB) in Fig. 2. SSATB can achieve a larger non-local receptive field. This hourglass-shaped structure further expands the receptive field.

**Blind Spot SR.** Taking $\boldsymbol{I}_0^{LR}$ as the training target, if all images are fed into CASR-DSAT, SR images are affected by noise. This is because the input contains training target. During the training process of minimizing self-supervised loss, the network cannot learn noise distribution from the training data but tends to perform equivalent mapping. Inspired by blind spot denoising [12], we replace $\boldsymbol{I}_0^{LR}$ with the image whose pixel values are all zeros during training. This technique effectively avoids the equivalent mapping. During testing, $\boldsymbol{I}_0^{LR}$ are directly fed into CASR-DSAT to fully utilize the complementary information from camera array images.

## 2. Ablation Study

Here, we provide a comparative analysis between the Single Transformer and our Dual Transformer.

**Single Transformer VS dual Transformer.** To validate the necessity of self-attention Transformer with different dimensions, we compare the single Transformer with the dual Transformer. The first type of single Transformer employs spatial self-attention Swin Transformer block in both the feature extraction and feature reconstruction modules. The second type of single Transformer employs channel self-attention Transformer block in both modules. They are trained for 30 epochs on the synthetic training dataset. The proposed dual Transformer achieves a PSNR gain of 2.66 dB over the first type of single Transformer and 0.4 dB over the second type on the synthetic DIV2K test dataset. For the first type of single Transformer, because the texture of features extracted by spatial self-attention Transformer is smooth, resembling resampling, which disrupts the aliasing patterns in the original LR image. Compared to the second type of single Transformer, dual Transformer employs spatial self-attention Transformer blocks in the feature reconstruction module, leveraging the long-range spatial dependency to further eliminate residual noise and smooth zipper edges.